\documentclass[aps,twocolumn,showpacs,groupedaddress]{revtex4}

\usepackage[dvips]{graphicx}
\usepackage{amssymb,latexsym}
\usepackage{psfrag}
\usepackage{wasysym}

\begin{document}
\title{Intermittent dry granular flow in a vertical pipe}
\author{Yann Bertho, Fr\'ed\'erique Giorgiutti-Dauphin\'e and
Jean-Pierre Hulin}
\affiliation{Laboratoire FAST, UMR 7608, B\^at.~502, Universit\'e
Paris XI, 91\,405 Orsay Cedex (France)}

\begin{abstract}
The intermittent compact flow of glass beads in a vertical glass
pipe of small diameter is studied experimentally by combining
particle fraction, pressure, and air and grain flow rates
measurements with a spatio-temporal analysis of the flow. At the
onset of the flow, a decompaction front is observed to propagate
from the bottom to the top of the tube at a velocity much larger
than that of the grains. The blockage front also propagates
upwards and at a still higher velocity. The decompaction induces a
decreasing pressure wave strongly amplified as it propagates
upwards towards the top of the tube. Pressure variations of
3000\,Pa or more are detected in this region while particle
fraction variations are of the order of 0.02. Grain velocities
during the flow period also increase strongly at the top of the
tube while the corresponding fraction of total time decreases. A
1D numerical model based on a simple relation between the
effective acceleration of the grains and the particle fraction
variations reproduces the amplification effect and provides
predictions for its dependence on the permeability of the packing.
\end{abstract}

\maketitle

\section{Introduction}
Dry particle flows driven by gravity (as in emptying hoppers or
silos) or by a gas current (in pneumatic transport
\cite{Laouar98}) are encountered in many industrial processes.
Modeling such flows represents a challenging fundamental problem
since they involve complex interactions of moving grains with the
surrounding air, bounding walls and other grains. Various flow
regimes may be observed in these systems according to the spatial
distribution of the particle concentration and velocity field.
Examples of such regimes include the free-fall of particles at
high velocities and low particle fractions, slow compact flows
with high particle fractions, and density waves in which compact
and dilute zones alternate \cite{Raafat96}. These particle flows
may be stationary (in this case, particles -- or density waves --
propagate at a constant mean velocity) or display oscillations,
stick-slip motions, intermittency or blockage effects. Such
non-stationarities have been reported in the simultaneous flow of
air and grains in vertical pipes
\cite{Leung78,Knowlton86,Aider99,Bertho02} and are encountered in
a variety of settings. They represent a significant problem in
many large-scale industrial facilities such as silos or pneumatic
transport systems: intermittent flows may induce very large,
potentially destructive, pressure variations.

Several approaches have been suggested to model theoretically
granular flows in pipes, each of them being applicable to only a
part of the flow regimes. Kinetic theories
\cite{Savage79,Jenkins83} are best suited when the particle
fraction is lower than 0.5 and collisional effects play an
important part. In this regime, the time between two successive
collisions is large compared to the duration of the collision. At
higher particle fractions, grains are in contact with their
neighbors a large fraction of the time and friction forces between
particles and walls become significant. Although attempts have
been made to adapt the kinetic theory to these frictional regimes
\cite{Azanza99}, different approaches have often been used. In
their study of compact moving bed flows, Chen {\it et al.}
\cite{Chen84} use the classical Janssen theory
\cite{Janssen95,Duran00b} to estimate friction forces on the walls
from the relation between the vertical component of the
stress-tensor in the grain packing and the horizontal normal
stress component on the walls.

The present work deals with intermittent compact flows of grains
in a vertical tube of small diameter. In this case, the particle
fraction is very close to that of a static random packing of
particles, and the solid friction between particles and walls is
always important. The objective of this study is to analyze the
spatio-temporal characteristics of the onset and blockage of such
intermittent flows. Of special interest is the relation between
air pressure, particle fraction and grain velocities and the
manner in which variations of these quantities are amplified as
they propagate along the flow tube.

A similar flow regime is encountered in the ``ticking hourglass"
experiment \cite{Wu93,Maloy94,LePennec95,LePennec96,Veje01}
corresponding to a periodic intermittent granular flow between two
closed glass containers connected by a short vertical
constriction. In this case, intermittency is believed to be
largely due to the build-up of a pressure difference induced by
the grain flow between the two containers (both because air is
dragged downwards and because the empty volume available for air
increases with time in the upper container and decreases in the
lower one). Once the back pressure is sufficiently large to halt
flow, it decays due to air flow through the grain packing in the
constriction until it is low enough to permit flow. In this
experiment, the key phenomena are localized in a small volume
close to the constriction; in particular, static arches of
particles are believed to build up right above it, where transient
fractures of the packing are also observed.

In the intermittent compact flow down a long vertical tube
considered in the present study, the flow intermittency phenomenon
is distributed over most of the height of the tube. This allows
one to perform simultaneously a large variety of measurements
(local pressures and particle fractions, air and grain flow rates,
\ldots) at several locations, and so characterized the variations
of the flow with height. High resolution spatio-temporal diagrams
of particle motions along the tube are also obtained and allow for
the characterization of propagation of the onset and blockage of
the grain flow along the tube length.

In section~\ref{exp}, we describe the experimental set-up and
procedure. Then, in section~\ref{Expresult}, the qualitative
characteristics of the compact intermittent flows are described,
and quantitative data and estimates of air pressure, particle
fraction and flow rate variations are presented. In
section~\ref{analysis}, we discuss appropriate volume and mass
conservation equations and their bearing on the experimental
results. The various forces acting on the grains and the manner in
which they may induce intermittency are discussed in
section~\ref{Model}. The equations of motion are then written and
solved numerically in order to model the spatial variations of
pressure, particle fraction and grain velocities.
\section{Experimental set-up and procedure}
\label{exp} 
\begin{figure}[t!]
\includegraphics[width=7.5cm]{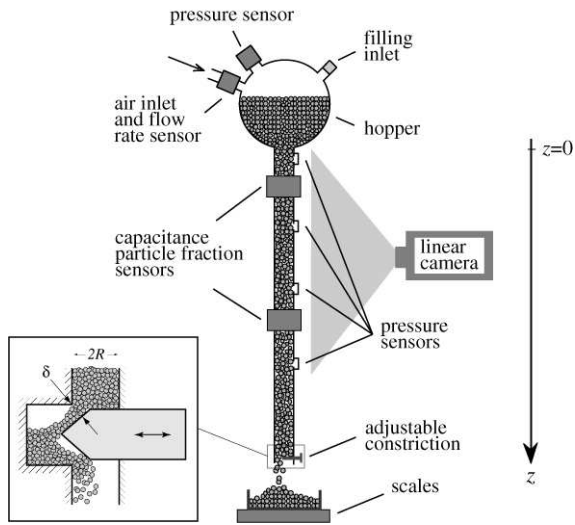}
\caption{View of the flow experiment. - Insert: sketch of the
constriction at the bottom end of the pipe.}
\label{fig:dessinmanip}
\end{figure}

The granular flow takes place in a vertical glass tube
of radius $R$\,=\,(1.5\,$\pm$\,0.025)\,mm and length
$L$\,=\,1.25\,m (Fig.~\ref{fig:dessinmanip}).
Grains flow from a spherical hopper welded to the top of the pipe.
The bottom end is fitted with a variable constriction allowing one
to adjust the outflow: this constriction is realized by a cylinder
with a conical tip which can be precisely moved across the pipe
(see insert of Fig.~\ref{fig:dessinmanip}). The grains are
spherical glass beads of diameter
2$a$\,=\,(175\,$\pm$\,25)\,$\mu$m and density
$\rho$\,=\,(2.50\,$\pm$\,0.02)\,10$^3$\,kg\,m$^{-3}$. Electronic
scales are placed below the tube and their output is transmitted
to a computer at the rate of one sample per second. The mass flow
rate $F_m$ of the grains is then determined from the time
variation of the measured mass. The grain flow rate will be
characterized by the mean superficial velocity $q$ representing
the volume flow rate of grains per unit area with:
\begin{equation}
q = {F_m\over{\rho\pi R^2}}. \label{defq}
\end{equation}
The hopper is connected to the tube but otherwise closed except
for an air inlet. The volume flow rate $Q_a^\mathrm{meas}$ of air
into the hopper is measured by an on-line sensor (the precision is
$\pm$\,1\% at the largest flow rates). An additional transducer
measures the air pressure $p_h$ inside the hopper. Continuity
requires that the inflow of air compensates for the volume of both
air and grains leaving through the vertical tube. Therefore if
$p_h$ remains constant with time, $Q_a^\mathrm{meas}$ is the sum
of the volume flow rates of air and of the grains in the
experimental tube. The superficial velocity $q_a$ of air in the
tube (volume flow rate per unit area) thus satisfies:
\begin{equation}
q_a = {Q_a^\mathrm{meas}\over{\pi R^2}} - q.
\label{defqa}
\end{equation}
Four pressure sensors are placed along the tube at vertical
distances 200, 450, 700 and 950\,mm below the outlet of the
hopper. They are connected to the inside of the tube through
0.5\,mm diameter holes in the tube wall. A fine-meshed grid is
stretched across the holes on the external wall to prevent grains
from entering the sensors. An independent experiment has been
realized to estimate the pressure drop across the grid and the
portholes during typical pressure transients encountered in the
intermittent flows. Due to the low dead volume of the transducer
(23\,$\mu$l), the corresponding error on the pressure measurement
is less than 10\,Pa for the fastest transients: this will be seen
below to be much smaller than the pressure to be measured.

Local mean particle fraction variations in tube sections are
measured by an electrical capacitance sensor using two shielded
3\,mm diameter cylindrical electrodes pressed against the outside
tube walls and facing each other. The sensor is connected to a
$GR1620$ capacitance bridge (10\,kHz measurement frequency) and a
lock-in detector with a 680\,$\mu$s time constant. The noise level
is of order $10^{-5}$\,pF, which corresponds to particle fraction
variations of order 10$^{-3}$.

In order to test the device, an air-grain interface was moved
through the sensor and the output was recorded as a function of
the location of the interface. This allowed one to determine that
the measurement is averaged over a slice of the sample of typical
thickness 2\,mm parallel to the tube axis. The variation of the DC
output with the particle fraction is assumed to be linear and the
probes are calibrated by comparing readings obtained with an empty
tube and with the same tube filled with a static bead packing.
This assumption of a linear variation is frequently used in the
literature: in the present case, it is particularly justified by
the small particle fraction variations during the experiment and
by the random, approximately isotropic structure of the packing.
The particle fraction $c_{max}$ of the static bead packing
obtained after filling the tube is assumed to be equal to 0.63:
this value corresponds to a {\em Random Close Packing} of spheres
and is frequently obtained experimentally in similar systems. A
direct measurement of $c_{max}$ has been realized but its
precision is limited by the small amount of beads in the tube. A
slightly lower value $c_{max}$\,=\,0.61\,$\pm$\,0.02 was obtained
(this difference may be due to the small diameter of the tube
increasing the influence of wall effects reducing locally the
particle fraction). In view of the large uncertainty on this
measurement, the value 0.63 is retained in the following.

The capacitance measurements can be performed at chosen heights by
moving the sensors along the tube. The instantaneous outputs of
four pressure and capacitance sensors are digitized simultaneously
at a sampling frequency of up to 1\,kHz by a Spectral Dynamics
$SD195$ signal analyzer. Time-averaged values of all pressure,
capacitance and flow rates sensors are also recorded but at 1\,s
intervals.

The dynamical properties of the flow are studied from
spatio-temporal diagrams constructed from appropriately assembling
the output of a digital linear CCD camera. Light intensity
variations along a vertical line precisely aligned with the tube
are recorded at sampling rates of 500 lines per second with a
resolution of up to 2048 pixels (the equivalent pixel size on the
tube ranges from 10\,$\mu$m to 500\,$\mu$m depending on the
magnification). One thus obtains $2D$ images in which the $x$-axis
represents time while the $z$-axis corresponds to distance along
the tube.
\section{Experimental results on the intermittent compact flow}
\label{Expresult} The flow regimes are controlled by adjusting the
constriction at the bottom of the experimental tube. The compact
flow regime is observed for narrow constrictions, at typical
superficial velocities $q$ between 0.01\,m\,s$^{-1}$ and
0.08\,m\,s$^{-1}$. Narrowing further the bottom constriction leads
to a blockage of the flow while density waves appear at higher
flow rates. The typical aperture $\delta$ of the constriction (see
Fig.~\ref{fig:dessinmanip} insert) in the compact regime ranges
from 1 to 2\,mm (or approximately from 6 to 12 bead diameters). In
all cases, the particle fraction in the compact regime does not
differ by more than 0.02 from the particle fraction $c_{max}$ in a
static packing.

Decreasing the relative humidity $H$ of the air down to 40\% or
using ``clean beads'' with a smooth surface, changes significantly
the behavior of the grain flow: in this case, a continuous flow of
grains is observed (its velocity is roughly constant with time and
of the order of a few centimeters per second). In return, for
``rougher beads'' or relative humidity $H\apprge$\,50\%, the flow
is intermittent and may remain blocked for a large fraction of the
time, particularly near the hopper (observations using a SEM
microscope display particles with a size of up to several
micrometers at the surface of the ``rough beads'' with spacings
increasing with the size from a few $\mu$m up to a few
10\,$\mu$m). Experiments reported here have been performed at a
relative humidity $H$\,=\,(55\,$\pm$\,5)\% for which the grain
flow is intermittent. At the outlet of the constriction, flow is
modulated but does not stop completely. The modulation may be
periodic, particularly at large flow rates. In the remainder of
the paper, we shall focus on this latter {\em intermittent compact
flow} regime.
\subsection{Spatio-temporal characteristics of the intermittent flow}
\begin{figure}[t!]
\includegraphics[width=8cm]{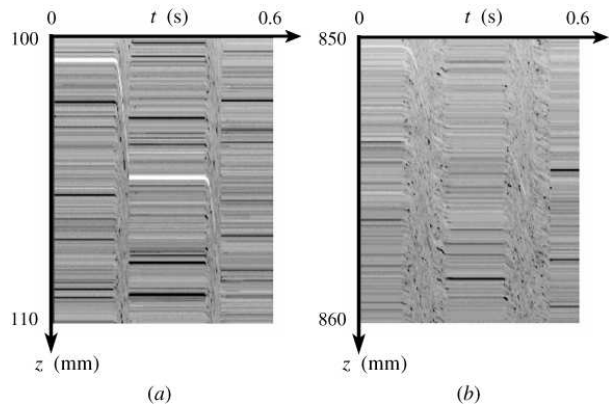}
\caption{Spatio-temporal diagram of the grain flow: ($a$)~at the
top of the tube (100\,mm to 110\,mm from the hopper), ($b$)~at the
bottom of the tube (850\,mm to 860\,mm from the hopper). Each
diagram corresponds to a time lapse of 0.6\,s and to a field of
view of $10$\,mm.} \label{fig:spatiodebut}
\end{figure}
Figure~\ref{fig:spatiodebut} displays two spatio-temporal diagrams
of the grain flow over two 10\,mm high segments located in the top
and bottom parts of the tube. Time corresponds to the horizontal
scale. The two diagrams were obtained at slightly different times
during a same experiment: due to the excellent periodicity of the
flow, they are however comparable. Time intervals during which
grains are at rest in the tube are marked by horizontal line
segments with constant grey levels; those during which grains are
moving down are marked by inclined striations: the slope of which
indicates the grain velocity. Starting from rest, the grains first
accelerate and reach a constant velocity before finally stopping
abruptly. Figures~\ref{fig:spatiodebut}($a$) and ($b$) show that
the fraction of the total time during which grains are flowing is
much larger at the bottom of the tube than at the top.

\begin{figure}[t!]
\includegraphics[width=8cm]{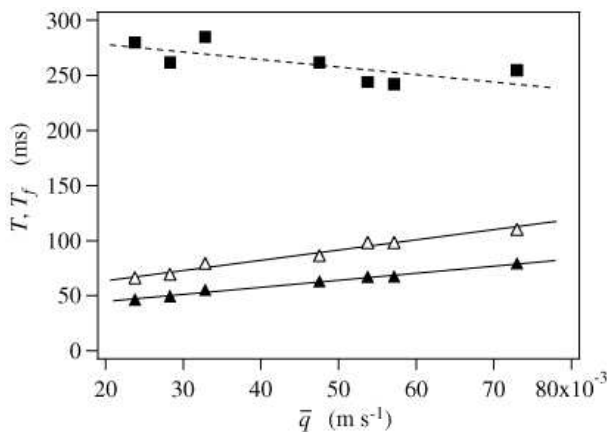}
\caption{Duration of flow phase $T_f$, as a function of the time
average $\overline{q}$ of the superficial velocity, at two
different distances $z$ from the top of the tube: $z$\,=\,250\,mm
($\blacktriangle$) and $z$\,=\,600\,mm ($\vartriangle$). Global
period of the intermittent flow $T$ ($\blacksquare$).}
\label{fig:dureeflowing}
\end{figure}

\begin{figure}[t!]
\includegraphics[width=7.5cm]{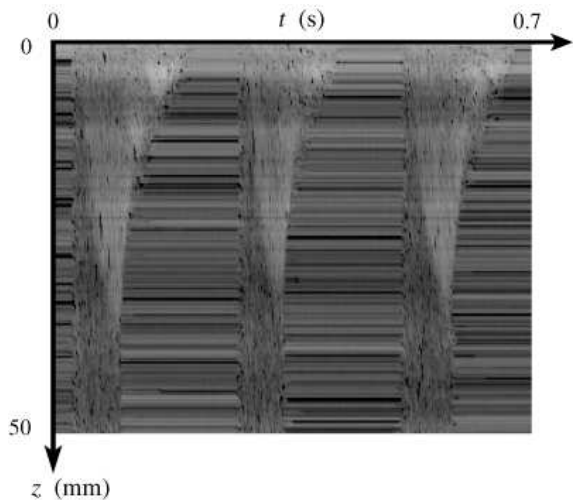}
\caption{Spatio-temporal diagram of the grain flow at distances
between 0\,mm and 50\,mm from the hopper at the top of the tube
($\overline q$\,=\,0.07\,m\,s$^{-1}$); total time lapse
corresponding to diagram: 0.7\,s.} \label{fig:spatiobulle}
\end{figure}

A quantitative comparison is made in Fig.~\ref{fig:dureeflowing}
in which the duration $T_f$ of the flow phase at two different
distances from the hopper and the period $T$ of the process are
displayed as a function of the time average $\overline {q}$ of the
superficial velocity ($T$ is the sum of the durations of the flow
($T_f$) and static phases). The time lapse $T_f$ is 40\% lower in
the upper section than in the lower one; it increases linearly
with $\overline q$ in both cases, while, on the contrary the
period $T$ decreases with $\overline q$.
Spatio-temporal diagrams also allow one to estimate qualitatively
the velocity of the grains in the flow phase from the slope of the
striations. Figures~\ref{fig:spatiodebut}($a$) and ($b$) show that
the grain speed is much higher at the top of the tube than at the
bottom as required by mass conservation. The particle fraction
$c$ varies little along the tube and the global displacement of the
particles during one period of the flow must be the same at all
heights. The mean value $\overline{v}$ of this velocity during the
flow phase can be estimated quantitatively from the relation:
\begin{equation}
\overline{v} = {{\overline q} \over {\overline c}}{T\over T_f},
\label{defvf}
\end{equation}
The variation of $\overline{v}$ with $\overline q$ in both
sections of interest is displayed in Fig.~\ref{fig:vitgrains(z)}:
\begin{figure}[t!]
\begin{psfrags}
\psfrag{v}{$\overline{v}$}
\includegraphics[width=8cm]{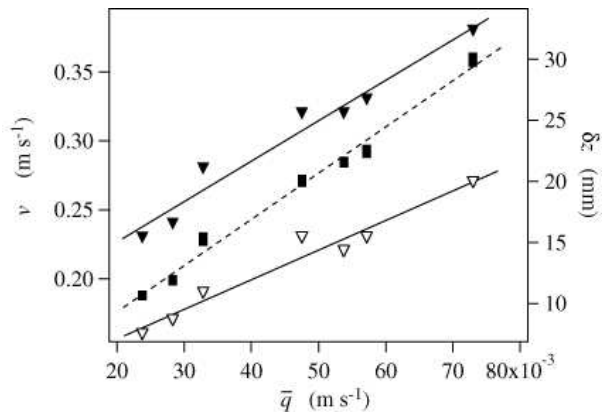}
\end{psfrags}
\caption{Mean grain velocity $\overline{v}$ vs. mean superficial
velocity $\overline q$ for
$z$\,=\,250\,mm ($\blacktriangledown$,~left axis) and
$z$\,=\,600\,mm ($\triangledown$,~left axis). Displacement $\delta
z$ of the grains during a flow period as a function of $\overline q$
($\blacksquare$,~right axis).} \label{fig:vitgrains(z)}
\end{figure}
in both cases, $\overline{v}$ increases roughly linearly with
$\overline q$. The displacement $\delta z$ of the grains during
one flow period is equal to $\overline{v}\times T_f$ and is also
plotted on Fig.~\ref{fig:vitgrains(z)}. For large mean flow rates,
the instantaneous grain flow rate $q$ in the tube exceeds the
maximum flow rate from the hopper. A low particle fraction {\em
bubble} thus builds up at the top of the tube: it appears as a
light zone in the spatio-temporal diagram of
Fig.~\ref{fig:spatiobulle} (the local particle fraction in this
bubble may be lower than 0.2).
The abrupt initial downwards slope of the contour of the light
zone reflects the fast downwards motion of the upper boundary of
the compact packing below the bubble. As soon as the motion of the
packing stops, the bubble gets filled up from the hopper, although
at a lower velocity that decreases with time (this may be due to
the build up of an adverse pressure gradient between the bubble
and the hopper).
\subsection{Spatial variations of the intermittency effect}
Information obtained from the spatio-temporal diagrams are
complemented by measurements of the local pressure variations.
\begin{figure}[t!]
\includegraphics[width=8cm]{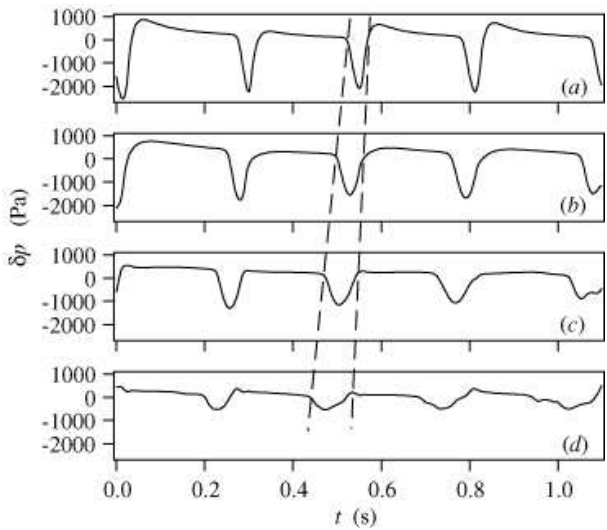}
\caption{Time recording of pressure variations $\delta p$ at four
different distances $z$ from hopper (from top to bottom,
$z$\,=\,200, 450, 700 and 950\,mm), for $\overline
q$\,=\,0.02\,m\,s$^{-1}$. Dashed lines correspond to the onset and
the blockage of the flow.} \label{fig:4presshaut}
\end{figure}
Figure~\ref{fig:4presshaut} displays time recordings of the
pressure variations $\delta p = p-p_o$ measured on four sensors at
different heights $z$ from the top ($p_o$ is the atmospheric
pressure). These curves are nearly periodic; the pressure drops
sharply while grains are flowing (by up to 3000\,Pa near the top
of the tube). Then the pressure increases back above $p_o$ and
decays slowly (or remains constant) while the grains are at rest.
The time average of the pressure is close to zero (at most 25\,Pa)
for all four transducers used for the measurements.

The flow has been filmed in the intermittent regime
using a 1000 frames per second high speed camera equipped with
analog input channels connected to the pressure transducers. The
onset of the flow at a given height is observed to coincide
exactly with the beginning of the pressure drop; the transition is
less sharp for the blockage since pressure diffusion is still
significant even after the grain flow has stopped. The region
between the two dashed lines in Fig.~\ref{fig:4presshaut}
corresponds to the flow phase (the lines are drawn, within
experimental error, between points corresponding to times at which
the flow is observed to start or stop). Comparing the widths of
this region in the different curves confirms that the fraction of
time corresponding to the flow phase increases with $z$. On the
contrary, the amplitude of the pressure drop decreases with $z$.
\begin{figure}[t!]
\includegraphics[width=7cm]{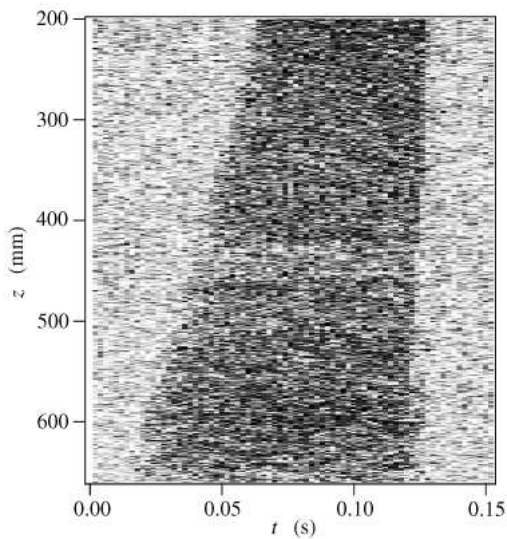}
\caption{Processed spatio-temporal diagram over a 500\,mm length
of tube (200\,mm\,$<$\,$z$\,$<$\,700\,mm) during one period of the
intermittent flow. Light zones correspond to the static phase and
the darker region to the flow.} \label{fig:spatiozoom}
\end{figure}

\begin{figure}[t!]
\begin{psfrags}
\psfrag{v}{{\small $|v_{df}|$,\ $|v_{bf}|$\ \ \ (m\,s$^{-1}$)}}
\includegraphics[width=8cm]{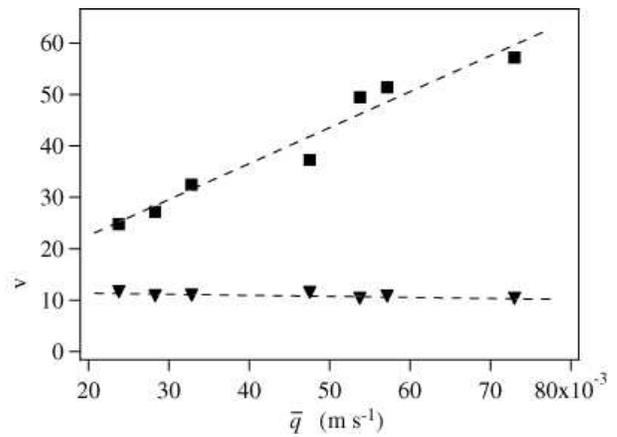}
\end{psfrags}
\caption{Variation of the absolute velocities of propagation of
the decompaction $|v_{df}|$ ($\blacktriangledown$) and blockage
$|v_{bf}|$ ($\blacksquare$) fronts with the mean superficial grain
velocity $\overline q$.} \label{fig:vitcompdecomp}
\end{figure}

The propagation of the onset and blockage of the flow were studied
more precisely on spatio-temporal diagrams of a 500\,mm high
region of the tube (Fig.~\ref{fig:spatiozoom}). The diagram has
been processed by subtracting each linear image from the next one
to detect more precisely the onset and blockage of the flow.
Domains corresponding to the static phase have a light shade (the
difference is not zero because of the noise of the camera and the
fluctuations of the illumination). The domain associated with the
flow phase is globally darker (the fluctuations of the grey levels
are large because one subtracts specular reflections of light on
the beads varying very much during the motion). The velocity
$|v_{df}|$ for the onset of the flow corresponds to the slope of
the left boundary in Fig.~\ref{fig:spatiozoom}: it was roughly
constant in that experiment but decreases slightly with distance
in others. The velocity $|v_{bf}|$ for the blockage is given by
the slope of the right boundary (averaged along the tube) and is
lower than $|v_{df}|$. Both the onset and the blockage propagate
upwards. The variations of $|v_{bf}|$ and $|v_{df}|$ with
$\overline q$ are plotted in Fig.~\ref{fig:vitcompdecomp}.
The velocity $|v_{bf}|$ for the blockage increases linearly with
$\overline q$ and ranges between 20 and 60\,m\,s$^{-1}$. This
variation will be discussed in section~\ref{volcons}. On the
contrary, the velocity $|v_{df}|$ of the location of the onset of
the grain flow (decompaction front) is independent of the mean
superficial velocity and is of the order of 11\,m\,s$^{-1}$. This
implies that the decompaction process is independent of the
downwards structure of the flow and of the width of the
constriction.

This velocity value may be related to the velocity of sound in
two-phase systems with a large density and compressibility
contrast between the two phases (water-air bubbles for instance).
In this case, by assuming an isentropic expansion of the gas, one
may compute the velocity of sound $v_\mathrm{sound}$ via
\cite{Wallis69}:
\begin{equation}
v_\mathrm{sound}^2={{\gamma p_{o}}\over {\rho c(1 - c)}},
\label{eq:vsound}
\end{equation}
in which $c$ and $\rho$ are the volume fraction and the density of
the dense phase, $p_o$ is the atmospheric pressure and $\gamma =
1.4$ for air which is considered as having negligible density.
This amounts to considering the air-grain system as a homogeneous
mixture of both high compressibility and density resulting in a
low sound velocity. Taking $\rho$ to be equal to the density of
the grains and $c$ to the mean particle fraction
$\overline{c}\simeq$\,0.615 leads to
$v_\mathrm{sound}\simeq$\,15\,m\,s$^{-1}$ which is indeed of the
order of the value $|v_{df}|\simeq$\,11\,m\,s$^{-1}$ deduced
experimentally ($0.615$ represents a typical value of the average
particle fraction as estimated from the measurements displayed in
Fig.~\ref{fig:cPQa} assuming as above that $c_{max}$\,=\,0.63).

\begin{figure}[t!]
\includegraphics[width=8cm]{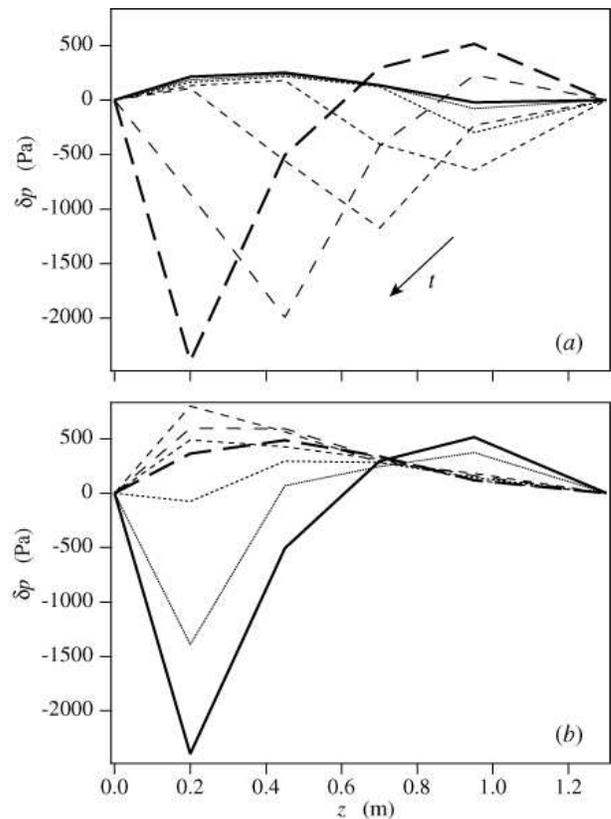}
\caption{Pressure deviation ($\delta p=p-p_o$) variations with
distance along the tube for several measurement times, for a mean
superficial velocity $\overline q$\,=\,0.02\,m\,s$^{-1}$.
($a$)~Solid line obtained during the static phase at $t$\,=\,0\,s.
Dashed lines are profiles obtained at various times during the
flowing phase: $t$\,=\,16\,ms, $t$\,=\,31\,ms, $t$\,=\,62\,ms,
$t$\,=\,86\,ms, $t$\,=\,109\,ms and $t$\,=\,133\,ms (dash length
and spacing increase with time). ($b$)~Solid line obtained near
the end of the flowing phase at $t$\,=\,133\,ms. Dashed lines are
profiles obtained at various times: $t$\,=\,141\,ms,
$t$\,=\,148\,ms, $t$\,=\,156\,ms, $t$\,=\,172\,ms,
$t$\,=\,211\,ms, $t$\,=\,266\,ms (dash length and spacing increase
with time).} \label{fig:gradP}
\end{figure}

Data points from the four curves in Fig.~\ref{fig:4presshaut}
together with the boundary conditions $p=p_o$ at both ends of the
tube provide the overall shape of the pressure profile along the
tube. The spatial variation of $\delta p=p-p_o$ at several
different times during one same period of variation of the flow
within the periodic sequence of Fig.~\ref{fig:4presshaut} is
displayed in Fig.~\ref{fig:gradP}.
The thick solid line in Fig.~\ref{fig:gradP}($a$) corresponds to
the static phase. A pressure maximum subsists in the central part
of the tube from the previous flow cycle and the associated
pressure gradients drive air towards both ends of the tube through
the motionless packing. When flow is initiated, a low pressure
zone appears at the bottom of the tube where grains are moving,
and propagates upwards. The amplitude of the pressure minimum
increases strongly during this propagation. This amplification
effect is discussed below in section~\ref{Model}. After the
pressure minimum has reached the top of the tube, pressure starts
to increase again [Fig.~\ref{fig:gradP}($b$)]. At first, the local
particle fraction reaches a minimum and the expansion of the gas
stops. Then, when the granular flow stops in the tube (first at
the bottom), the particle fraction reverts back to its static
value. Thus, air gets compressed above the atmospheric pressure
since some additional air has leaked between the grains during the
flow period (see for instance the curve corresponding to
$t$\,=\,172\,ms in Fig.~\ref{fig:gradP}($b$)). Finally, the
pressure profile along the static packing relaxes back to its
initial distribution at the beginning of the flow phase and a new
flow cycle may start again.

Note that when the intermittent flow regime does not appear
spontaneously, it can also be triggered externally by periodic
injections and suctions of air close to the constriction at a
frequency comparable to that of the natural oscillations. The
other characteristics of the flow such as the amplification of
pressure and particle fraction variations as they propagate
upwards are the same as in the spontaneous intermittent flow
regime.
\subsection{Particle fraction and flow rate variations}
\begin{figure}[t!]
\includegraphics[width=8cm]{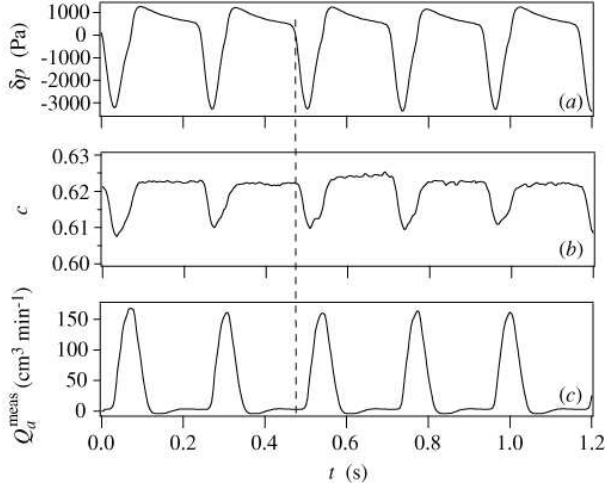}
\caption{Time recordings of the pressure deviations $\delta p$ and
particle fraction $c$ variations (at $z$\,=\,200\,mm) and of the
measured air flow-rate $Q_a^\mathrm{meas}$ into the hopper.
$Q_a^\mathrm{meas}$ is defined as positive for air flowing
downwards. The dashed line corresponds to the onset of the flow.}
\label{fig:cPQa}
\end{figure}
Spatio-temporal diagrams have allowed us to identify granular
motions in the various phases of the intermittency process; their
relation with the spatial distribution of the pressure $\delta p$
has also been analyzed. We proceed by discussing flow rate and
particle fraction measurements which provide additional
information on the flow structure.

The variations of the particle fraction $c$ and pressure $\delta
p$ at a fixed height, and of the flux $Q_a^\mathrm{meas}$ of air
into the hopper are displayed in Fig.~\ref{fig:cPQa}. Variations
of the local particle fraction $c$ associated with the onset and
blockage of the flow are only of the order of 0.015 in the present
example and were never higher than 0.02. Moreover, these
variations are evidently synchronous with those of the local
pressure $p$ in the flowing phase. Note the slow pressure decay
when the grains are at rest. The amplitudes of the variations of
$c$ and $\delta p$ decrease, while their duration increases with
distance from the hopper. In addition to demonstrate this close
correlation between the variations of $p$ and $c$, our study
indicates an increase of the amplitude of the $c$ variations with
the instantaneous grain velocities.

The time recording of $Q_a^\mathrm{meas}$ indicates a strong
increase of the downwards air flow rate during the flow phase.
This increase reflects the entrainment of air by the granular flow
and the compensation of the volume of grains leaving the hopper by
the inflow of air. The variation of $Q_a^\mathrm{meas}$ is
slightly delayed relative to those of $p$ and $c$ since a time is
required for the grain flow to propagate to the top of the tube
and into the hopper. A transient upward gas flow
($Q_a^\mathrm{meas}<$\,0) is observed after grains stop moving in
the upper part of the tube, corresponding to air being expelled
through the grain packing.
\section{Quantitative analysis of the intermittent flow components}
\label{analysis}
\subsection{Conservation equations for air and grains}
\label{volcons} In the following, $z$\,=\,0 corresponds to the top
of the tube and the coordinate $z$ increases downwards. Downflow
is defined as being positive. $v(z,t)$ and $u(z,t)$ are,
respectively, the grain and air velocities (averaged over the flow
section) in the laboratory frame at a time $t$ and a distance $z$
from the hopper. They are related to the superficial velocities
$q(z,t)$ and $q_a(z,t)$ and to the local particle fraction
$c(z,t)$ (also averaged over a flow section) by the relations:
\begin{equation}
q = c\,v,
\label{eq:q(z,t)}
\end{equation}
\begin{equation}
q_a = (1-c)\,u.
\label{eq:qair(z,t)}
\end{equation}
The superficial velocity $q$ of the grains is related to the
variation of $c$ with time by the volume conservation equation:
\begin{equation}
{{\partial c}\over {\partial t}} = - {{\partial q}\over {\partial
z}} = - c {{\partial v}\over {\partial z}} - v {{\partial c}\over
{\partial z}}.
\label{eq:consgrain}
\end{equation}
A similar relation can be written for the mass flux of air $\rho_a
q_a$ ($\rho_a$ being its density):
\begin{equation}
{\partial \over {\partial t}}[\rho_a (1 - c)] = - {\partial \over
{\partial z}} (\rho_a q_a). \label{eq:consair1}
\end{equation}
Assume that the variations of the gas pressure are isentropic so
that:
\begin{equation}
{p\over {{\rho_a}^{\gamma}}}={p_o\over {{\rho_{a_o}}^{\gamma}}} =
\mathrm{constant}, \label{eq:adiabatic}
\end{equation}
($p_o$ is the atmospheric pressure, $\rho_{a_o}$ the air density
at $p_o$ and $\gamma$=1.4). Then Eq.~(\ref{eq:consair1}) becomes:
\begin{equation}
{{\partial c}\over{\partial t}} = {{(1-c)}\over {\gamma p}}
{{\partial p}\over{\partial t}} + {{\partial q_a}\over {\partial
z}} + {{q_a}\over{\gamma p}} {{\partial p}\over {\partial z}}.
\label{eq:consair3}
\end{equation}
A first prediction provided by mass conservation conditions is the
velocity $v_{bf}$ at which the blockage front moves up the tube.
We define $c_{max}$ to be the particle fraction in the static
column and $c^+$ and $v^+$ the particle fraction and grain
velocity right above the front. $v_{bf}$ is obtained by expressing
the conservation of mass of the grains in a slice of thickness
$\delta z$ while the front moves from the bottom to the top of the
slice:
\begin{equation}
|v_{bf}| = {{v^+\, c^+}\over{c_{max} - c^+}}. \label{eq:vbf}
\end{equation}
Since $v^+$ increases linearly with the superficial velocity $q$,
equation~(\ref{eq:vbf}) implies therefore that $|v_{bf}|$ should
also increase linearly with $q$. This agrees with the trend
displayed in Fig.~\ref{fig:vitcompdecomp}. Using the experimental
values of $v^+$, $c^+$ and $c_{max}$ provides an order of
magnitude of $v_{bf}$ ($|v_{bf}|\simeq$\,20\,m\,s$^{-1}$) which is
in a good agreement with that obtained in
Fig.~\ref{fig:vitcompdecomp}.
\subsection{Relative magnitude of air flow components}
Another issue regarding the flow of air through the system is the
relative importance of the passive drag of air by the moving
grains and of the permeation of air through the grain packing. The
first flow component is equal to $q(1-c)/c$ and the second is
assumed to be related to the pressure gradient by Darcy's
equation:
\begin{equation}
q_a^\mathrm{Darcy}=-{K\over \eta _a}{\partial p\over \partial z},
\label{eq:darcylaw}
\end{equation}
where $\eta_a$ is the viscosity of air and $K$ the permeability of
the grain packing. The superficial velocity $q_a$ of air is then
given by:
\begin{equation}
q_a=-{K\over \eta _a}{{\partial p}\over {\partial z}}+{q\over
c}(1-c).
\label{eq:qadarct}
\end{equation}
Equations~(\ref{eq:consgrain}), (\ref{eq:consair3}) and
(\ref{eq:qadarct}) can be combined to yield:
\begin{equation}
{{\partial p} \over {\partial t}}= - v{{\partial p}\over{\partial
z}} - {{\gamma p}\over {(1 - c)}} {{\partial v} \over {\partial
z}} + D {{\partial^2 p} \over {\partial z^2}},
\label{eq:airconservation3}
\end{equation}
in which the coefficient $D$ satisfies:
\begin{equation}
D = {{\gamma pK}\over{(1-c)\eta_a}}, \label{eq:coefdif}
\end{equation}
(the second order term $[D/(\gamma p)] (\partial p/\partial z)^{2}$
has been neglected in Eq.~(\ref{eq:airconservation3})).
Considering locally the system as a packing of spherical beads of
diameter $d$, the permeability $K$ of the grains can be estimated
from the Carman--Kozeny relation:
\begin{equation}
K={(1-c)^3\,d^2\over{180\,c^2}}.
\label{eq:carman}
\end{equation}
In the following, we assume that to leading order, $p= p_o$ and $c
= {\overline c}$, so that $D$ and $K$ are constants.

\begin{figure}
\includegraphics[width=8cm]{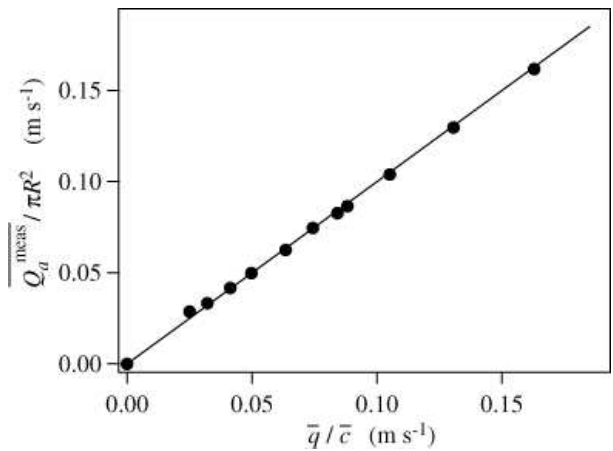}
\caption{Average inflow of air as a function of $\overline q /
\overline c$.} \label{fig:debitsairgrains}
\end{figure}

The average of $q_a^\mathrm{Darcy}$ over a flow period is
determined by plotting the time averaged value of
$Q_a^\mathrm{meas}$ as a function of the ratio ${\overline q}
/{\overline c}$ in Fig.~\ref{fig:debitsairgrains} (this amounts to
assuming that the mean grain velocity is equal to ${\overline q}
/{\overline c}$ which is approximately verified due to the small
relative variations of $c$ in these experiments).
Volume conservation implies that
$\overline{Q_a^\mathrm{meas}}/{\pi R^2}$ and the sum $\overline q
+ \overline{q_a}$ of the volume flow rates of air and grains are
equal [see Eq.~(\ref{defqa})]. Using Eq.~(\ref{eq:qadarct}), this
leads to:
\begin{equation}
\overline{Q_a^\mathrm{meas}\over {\pi R^2}} = \overline q +
\overline{q_a} = \overline{q_a^\mathrm{Darcy}} + {\overline q\over
\overline c}.
\label{eq:debtot}
\end{equation}
Experimentally, the variation of $\overline{Q_a^\mathrm{meas}}$
with $\overline q /\overline c$ is linear with a slope equal to
1\,$\pm$\,0.02 in Fig.~\ref{fig:debitsairgrains}. These results
indicate that the time average of $q_a^\mathrm{Darcy}$ is
negligible within experimental error although its instantaneous
value is non zero. The relative motion of air and of the grains is
therefore significant only during transient flow.
\subsection{Pressure variation mechanisms}
\begin{figure}[t!]
\includegraphics[width=8cm]{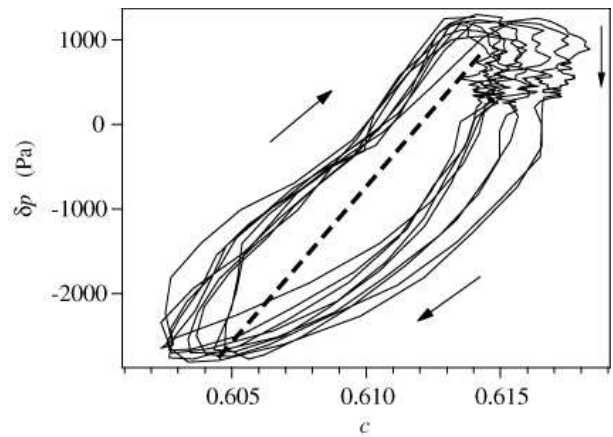}
\caption{Pressure vs. particle fraction variations during 10
cycles of the intermittent flow. Arrows indicate the direction of
the variation on each part of the curve: the vertical phase
corresponds to the slow pressure relaxation through the motionless
grain packing, the lower section corresponds to the onset of the
flow and the upper one to the blockage of the flow. The dashed
line has the theoretical slope $\gamma p_o /(1-\overline{c})$.}
\label{fig:presscomp}
\end{figure}

The dynamical properties of the system are largely determined by
spatial and temporal variations of the air pressure. We first
assume that variations of density $\rho_a$ are only due to those
of the local particle fraction $c$. This amounts to neglecting the
permeation of air relative to the grains. Then, ${\partial
q_a}/{\partial z}$ and $[q_a/(\gamma p)]\,\partial p/\partial z$
are equal to zero in which case Eq.~(\ref{eq:consair3}) becomes:
\begin{equation}
{1\over (1-\overline{c})}{{\partial c}\over {\partial t}} =
{1\over {\gamma p_o}} {{\partial p}\over {\partial t}}.
\label{eq:consred}
\end{equation}
($p$ and $c$ are replaced by $p_o$ and $\overline c$ within a
first order approximation). This relation is tested in
Fig.~\ref{fig:presscomp} by plotting pressure variations as a
function of particle fraction variations: for variations small
compared to the mean value, data points should be located on the
dashed line of slope $\gamma p_o/(1-\overline{c})$ shown on
Fig.~\ref{fig:presscomp}.
The two parameters are clearly strongly correlated and the slope
of the curves has the right order of magnitude in the initial and
final phases of the variation. However, pressure is higher in the
blockage phase than during the development of the flow, indicating
that ${\partial q_a}/{\partial z}$ cannot be neglected in
Eq.~(\ref{eq:consair3}) (the last term $[q_a/(\gamma p)]\,\partial
p/\partial z$ is of lower order and should remain negligible).
Some air gets sucked into the lower pressure regions when grains
flow; this leads to an overpressure when flow stops and the
particle fraction increases.

This overpressure decays slowly during the remainder of the static
phase (short vertical part of the curve at top right) due to the
finite permeability of the packing. In this static phase of the
flow sequence, $v=0$. The two first terms on the right hand side
of Eq.~(\ref{eq:airconservation3}) are equal to zero leading to
the diffusion equation:
\begin{equation}
{{\partial p }\over{\partial t}} = D {{\partial^2p}\over{\partial z^2}}.
\label{eq:eqdif}
\end{equation}
Experimentally, the particle fraction varies from 0.605 to 0.625;
the permeability $K$ computed from Eq.~(\ref{eq:carman}) ranges
therefore from 23\,10$^{-12}$ to 29\,10$^{-12}$\,m$^{-2}$. Taking
$\eta_a$\,=\,1.85\,10$^{-5}$\,m$^2$\,s$^{-1}$,
$p_o$\,=\,10$^5$\,Pa and a mean value of $K$
($K$\,=\,26\,10$^{-12}$\,m$^{-2}$), one obtains
$D$\,$\simeq$\,0.5\,m$^2$\,s$^{-1}$. The characteristic diffusion
distance during the period $T\,\simeq$\,0.25\,s of the
intermittent flow is $\sqrt{DT}\,\simeq$\,0.35\,m which is
comparable to the length over which pressure fluctuations diffuse.
During the onset of the flow which takes place over a shorter time
(a few hundredths of a second) the diffusion of the pressure
variations is sufficiently small for Eq.~(\ref{eq:consred}) to be
approximately valid. One thus concludes that the permeability of
the grain packing is high enough to allow for relative motions
between air and the grains during the flow phase; but too low to
allow for a global mean flow of air through the packing. In the
previous computation, the initial particle fraction of the static
packing is taken equal to $c_{max}$\,=\,0.63. Assume that wall
effects reduce this value to 0.61 as discussed in
section~\ref{exp} so that the particle fraction varies from 0.585
to 0.605 during a flow cycle. The mean corresponding value of $K$
is \,32\,10$^{-12}$\,m$^{-2}$ so that $D$ would increase to
$\simeq$\,0.6\,m$^2$\,s$^{-1}$ and remain of the same order of
magnitude as for $c_{max}$\,=\,0.63.

Our experiments suggest that the intermittent flows result from
the amplification of pressure and velocity fluctuations appearing
spontaneously or induced in the constriction at the bottom of the
tube. One expects the period of the intermittency to be determined
by the characteristics of the flow over the full length of the
tube and not just at the constriction. The period of the flow is
of the order of the characteristic diffusion time of air over a
distance of half of the tube length. This diffusion time
represents the order of magnitude of the time necessary for
pressure perturbations induced during the flow period to relax
before flow is reinitiated. It may thus be closely related to the
duration of the static phase of the periodic flow.
\section{Modeling of the intermittent flow}
\label{Model} In this section, the amplification of the
decompaction wave as it propagates towards the top of the flow
tube is modeled numerically. This approach yields insight into the
dependence of the amplification on the permeability of the grain
packing.
\subsection{Equations of motion and force distributions}
The equations of motion and force distribution in the granular
packing will first be written. They will then be solved
numerically together with the conservation equations for air and
grains. For simplicity, one models the case of a lower particle
fraction zone propagating upwards in a vertical channel.

The equation of motion reflecting momentum conservation for both
air and grains contained in a vertical slice with tube radius $R$
can be written as:
\begin{equation}
\rho c {dv \over dt} = \rho cg - \frac{\partial p}{\partial z} -
{{\partial \sigma_{zz}} \over {\partial z}} - {2 \over
R}{\sigma_{zr}}, \label{eq:mvtpart}
\end{equation}
(the equation has been divided by the section $S=\pi R^2$).
Acceleration terms for air have been neglected due to the large
density difference between the air and grains. The friction forces
on the air applied at the walls are similarly negligible due to
their low area compared to that of the grains.

Note that when Eq.~(\ref{eq:mvtpart}) (excluding the gravity,
$\sigma_{zz}$, and $\sigma_{zr}$ terms) is combined with
Eq.~(\ref{eq:airconservation3}) (neglecting the second order
$v\,\partial p/\partial z$ and diffusion terms), one obtains a
sound wave propagation equation with a velocity given by
Eq.~(\ref{eq:vsound}).

The left side of Eq.~(\ref{eq:mvtpart}) corresponds to the
acceleration term and the first term on the right to the weight of
the grains. $\sigma_{zr}$ and $\sigma_{zz}$ represent,
respectively, friction forces on the grains at the lateral walls
and vertical stress forces between grains in the tube section;
they are assumed to be constant around the perimeter of the tube
and in each section. $\sigma_{zr}$ is then estimated by applying
Janssen's model \cite{Janssen95,Gennes99,Duran00b} assuming first
that $\sigma_{zr} = \mu \sigma_{rr}$ following Coulomb's relation.
$\sigma_{rr}$ is taken to be $K_J \sigma_{zz}$, where the
coefficient $K_J$ depends on the structure of the packing and of
the surface properties of the grains. Equation (\ref{eq:mvtpart})
then becomes:
\begin{equation}
\rho c {{\partial v} \over {\partial t}} + \rho c v{{\partial v}
\over {\partial z}} = \rho c g -{{\partial p} \over {\partial z}}-
{{\partial {\sigma_{zz}}} \over {\partial z}}- {\sigma _{zz} \over
\lambda}, \label{eq:mvtpartjan}
\end{equation}
in which the characteristic length $\lambda$ is given by:
\begin{equation}
\lambda = {R \over 2\mu K_J}.
\end{equation}
In a static packing, all terms involving the grain velocity
vanish: the weight of the grains is balanced by the pressure
gradient and the stress forces at the walls and in the tube
section. When flow starts, the local particle fraction decreases
[Fig.~\ref{fig:cPQa}($b$)] which, in turn, reduces friction forces
at the walls, resulting in an acceleration of the flow. No exact
relation is available on the variations of the stress forces.
It will be assumed in the following that the
resultant of the weight of the grains, the pressure gradient and
the friction forces increases linearly with the deviation of the
particle fraction $c$ from its static value $c_{max}$.
Equation~(\ref{eq:mvtpartjan}) then becomes:
\begin{equation}
{{\partial v} \over {\partial t}} = -v{{\partial v}\over {\partial
z}} + J(c_{max}-c)\left (g-\frac{1}{\rho c}\frac{\partial
p}{\partial z}\right ), \label{eq:mvtpart2}
\end{equation}
where $J$ is the parameter characterizing the magnitude of the
variation. This assumption represents a first order approximation:
it will be valid in the initial phase of the development of the
perturbation as long as the derivative of the resultant force with
respect to the particle fraction is non zero for $c = c_{max}$.
Other types of functions may be envisioned to replace
$J(c_{max}-c)$: many will also reproduce an amplification
phenomenon provided they predict a resultant force in the right
direction with a monotonous dependence on ($c_{max}-c$) and if (as
will be seen below) the prefactor equivalent to $J$ is large
enough.
\subsection{Numerical procedure}
Numerical simulations are realized on a simplified 1D model in
which a region of reduced particle fraction (called \emph{bubble})
moves upwards inside a vertical tube filled with an otherwise
 grain packing. The bubble length is much smaller than that
of the tube and it is assumed to be far enough from either of the
ends that boundary conditions do not influence the motion. As
suggested by the experimental observations, the particle fraction
and the velocity are discontinuous at the bottom end of the bubble
(the blockage front), while they vary continuously at the top (the
decompaction region). Air pressure is continuous in both cases but
the pressure gradient is discontinuous at the lower front.

In order to solve the problem numerically, one uses the
conservation equations for grains~(\ref{eq:consgrain}) and air
(\ref{eq:consair3}) established in section~\ref{analysis} and the
equation of motion~(\ref{eq:mvtpart2}) of the grains. For the sake
of practicality, Eq.~(\ref{eq:airconservation3}), in which Darcy's
law is used to determine $q_a$, replaces Eq.~(\ref{eq:consair3}).
In the static regions of the packing, all terms in
Eqs.~(\ref{eq:consgrain}) and (\ref{eq:mvtpart2}) are equal to
zero. Equation~(\ref{eq:airconservation3}) reduces then to the
diffusion equation~(\ref{eq:eqdif}) for air with a coefficient $D$
given by Eq.~(\ref{eq:coefdif}).

At large distances below and above the bubble, the air pressure
$p$ is equal to the atmospheric pressure $p_o$; $c$ and $v$ are
respectively equal to $c_{max}$ and zero both far above the bubble
and everywhere below the blockage front. Additional boundary
conditions corresponding to the mass conservation for air and for
the grains must also be verified at the lower front (in addition
to pressure continuity). In the reference frame moving with the
front at $v_{bf}$, grain mass conservation is expressed by
Eq.~(\ref{eq:vbf}) which can be rewritten as:
\begin{equation}
(v^+-v_{bf})c^+=-v_{bf}c_{max}. \label{eq:grainconservafront}
\end{equation}
In the same way, the mass conservation of air implies:
\begin{equation}
(v^+-v_{bf})(1-c^+) - {K\over {\eta_a}} \left. {\partial p} \over
{\partial z}\right |_+ = -v_{bf}(1-c_{max}) - {K\over {\eta_a}}
\left. {\partial p} \over {\partial z}\right |_-.
\label{eq:airconservfront}
\end{equation}
The indexes $^+$ and $^-$ correspond to values close to the
blockage front respectively inside the bubble and in the static
zone. Combining Eqs.~(\ref{eq:grainconservafront}) and
(\ref{eq:airconservfront}) leads to:
\begin{equation}
v^+ - {K\over {\eta_a}} \left. {{\partial p}\over {\partial
z}}\right |_+ = -{K\over {\eta_a}} \left. {{\partial p}\over
{\partial z}}\right |_-.
\label{eq:airconservfront2}
\end{equation}
The initial particle fraction, velocity and pressure profiles are
assumed to follow a $1/\cosh (z)$ variation with a cutoff at the
lower front ($z = z_{f}$) for $c$ and $v$. Pressure is continuous
at $z = z_{f}$ with a faster decrease below. The corresponding
profiles are displayed in Figs.~\ref{fig:compsimul}($a$--$c$)
(rightmost solid curves). The initial particle fraction and
pressure variations are related by Eq.~(\ref{eq:adiabatic}) above
the front (this amounts to assuming that the bubble has been
created through a local decompaction of the grains inducing an
isentropic adiabatic expansion of the air). The amplitudes of
these initial variations are respectively $\delta c$\,=\,0.001,
$\delta p$\,=\,600\,Pa and $\delta
v$\,=\,3\,10$^{-3}$\,m\,s$^{-1}$.

Pressure, particle fraction and grain velocity profiles at
subsequent time steps are then computed by solving numerically
Eqs.~(\ref{eq:consgrain}), (\ref{eq:airconservation3}) and
(\ref{eq:mvtpart2}) in the reference frame of the front using a
finite difference Euler explicit scheme. For the resolution of
Eq.~(\ref{eq:mvtpart2}), an upwind scheme was required to avoid
numerical instabilities due to the advection term $v\,\partial
v/\partial z$. The global length of the simulation domain is
1.4\,m and the mesh size 2\,mm. The origin of distances is taken
at the initial location $z_f (0)$ of the front (the vertical
dashed line on Figs.~\ref{fig:compsimul}($a$--$c$)). Simulations
correspond to a time lapse of 27\,ms during which the bubble
propagates over a distance of 0.4\,m. At each step of the
simulation, the pressure, the particle fraction and the velocity
are determined at all distances from the front and the front
velocity is also obtained. This allows one in particular to
determine the new location of the front after each time step and
to translate distances from the front into distances in fixed
coordinates which will be used in subsequent plots.
\subsection{Numerical results}
\subsubsection{Qualitative observations}
\begin{figure}[ht!]
\begin{psfrags}
\psfrag{v}{\small $v$}
\includegraphics[width=8cm]{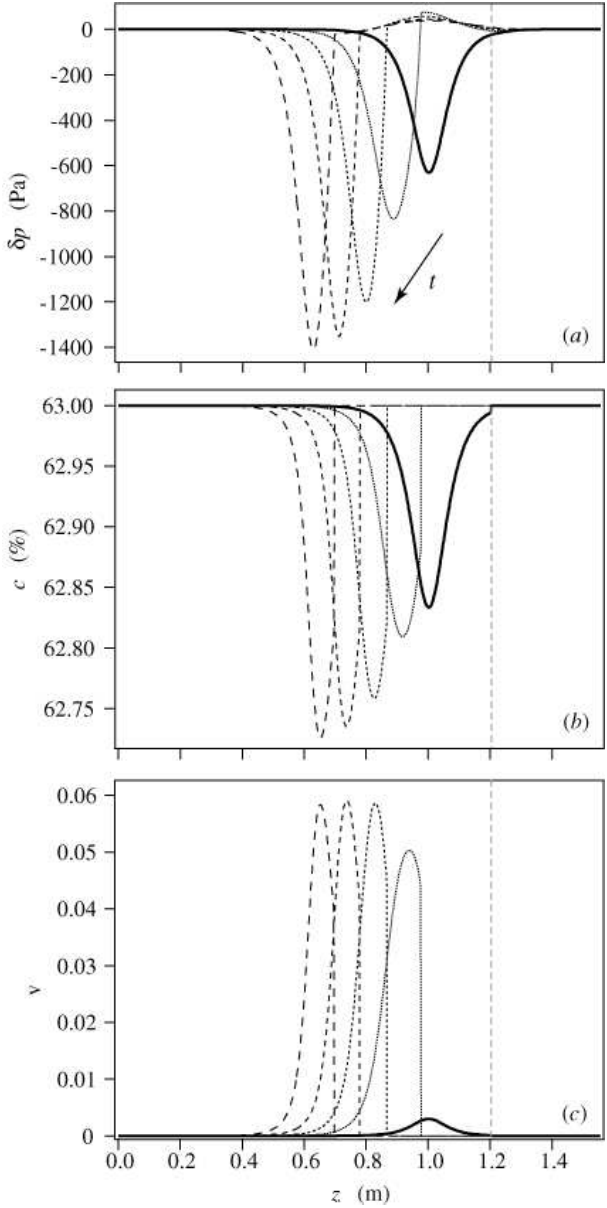}
\end{psfrags}
\caption{Numerical pressure~($a$), particle fraction~($b$) and
velocity~($c$) profiles, obtained for $J$\,=\,450,
$K$\,=\,20\,10$^{-12}$\,m$^2$ and $c_{max} = 0.63$. From right to
left: solid lines: initial profile at $t$\,=\,0\,s - dashed lines:
$t$\,=\,9\,ms, $t$\,=\,15\,ms, $t$\,=\,21\,ms, $t$\,=\,27\,ms
(dash length and spacing increases with time). The vertical dashed
lines represents the position of the front $z_f$ at $t$\,=\,0\,s.}
\label{fig:compsimul}
\end{figure}

Particle fraction, velocity and pressure profiles at different
times obtained for a typical set of values of $J$, $K$ and
$c_{max}$ are displayed in Fig.~\ref{fig:compsimul}. The other
parameters of the simulation are given by the experimental values
listed above or computed from $J$ and $K$.
A clear amplification effect is observed for all three parameters.
The velocity $v$ reaches its maximum value faster than both the
particle fraction $c$ and the pressure $\delta p$ (typically after
0.3\,m versus 0.5\,m). The ratio of the maximum and initial
amplitudes is also higher (20 for $v$ versus 2 and 1.6 for $\delta
p$ and $c$). A small localized overpressure appears in the static
region close to the initial location of the bubble: it is due to
air advected by the grains into this region at short times [from
Eq.~(\ref{eq:airconservfront2})]. At later times, the pressure
gradient close to the front increases so that the diffusion of air
back into the bubble balances the advection term. The increase of
the amplitude of the variations of $\delta p$ and $c$ is
consistent with the decrease of the width of these curves: the
total area under these curves must remain constant as required by
mass conservation of air and grains. The velocity $v_{bf}$ of the
lower front is also roughly constant with a value
$|v_{bf}|\simeq$\,12.5\,m\,s$^{-1}$ of the same order of magnitude
as that observed in our experiments.

While the computed amplitude of the pressure variations is of
comparable magnitude to the experimental ones, particle fraction
and velocity variations are smaller. This is likely a result of
the difference between the model case of a localized bubble and
the actual intermittent flow which sometimes extends over the
entire length of the tube.
\subsubsection{Quantitative dependence on the modelization parameters}
\begin{figure}[t!]
\begin{psfrags}
\psfrag{v}{\small $|v_{bf}|$}
\includegraphics[width=8cm]{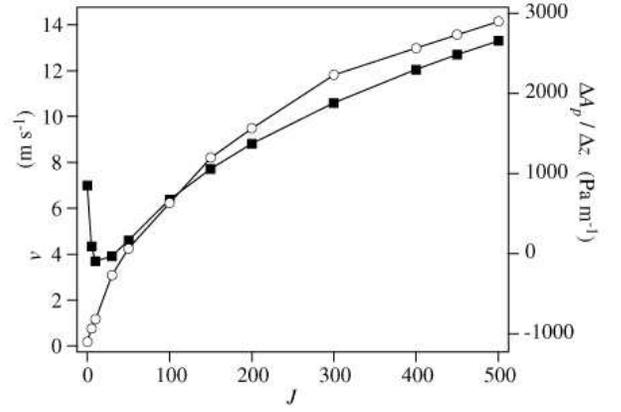}
\end{psfrags}
\caption{Variation of the amplification parameter $\Delta
A_p/\Delta z$ ($\circ$) and of the absolute front velocity
$|v_{bf}|$ ($\blacksquare$) with the force parameter $J$, for
$K$\,=\,20\,10$^{-12}$\,m$^2$.} \label{fig:Jinfluence}
\end{figure}

The key adjustable parameter in the model is the coefficient $J$
in Eq.~(\ref{eq:mvtpart2}) (that characterizes the effective
acceleration of the grains when the particle fraction $c$
decreases below $c_{max}$). Figure~\ref{fig:Jinfluence} displays
the dependance on $J$ of the front velocity $|v_{bf}|$ and of the
parameter $\Delta A_p/\Delta z$ characterizing the amplification.
$\Delta A_p/\Delta z$ is determined by plotting the minimum value
$A_p$ of the pressure variation curves [as shown in
Fig.~\ref{fig:compsimul}($a$)] as a function of the coordinate
$z_f$ of the front for each curve. After a short transient, $A_p$
increases quickly with $z_f$ and the variation levels off
thereafter. The amplification parameter $\Delta A_p/\Delta z$ is
taken equal to the maximum absolute value of the slope
$|dA_p/dz_f|$ of this curve. The permeability $K$ of the grain
packing has been taken equal to the value
$K$\,=\,20\,10$^{-12}$\,m$^2$). Using the values of ($c_{max}-c$)
deduced from the simulations, the maximum value $J$\,=\,500 in
Fig.~\ref{fig:Jinfluence} corresponds in Eq.~(\ref{eq:mvtpart2})
to an effective acceleration of the grains of the order of $g/3$
or lower.

For $J$\,=\,0, $\Delta A_p/\Delta z$ is negative and initial
variations of $\delta p$ and $c$ damp out as the bubbles propagate
upwards. $\Delta A_p/\Delta z$ becomes equal to 0 for $J$\,=\,40
and increases steadily at higher values. The front velocity
$|v_{bf}|$ also increases with $J$ in the amplification domain but
never becomes zero. Its value is of the same order of magnitude as
the experimental front velocities, particularly when $J$ is large.
We note that in the case of a bubble of slowly varying length,
$v_{df}$ and $v_{bf}$ are nearly equal while they differ in the
experiments so that more detailed quantitative comparisons would
be irrelevant.
These results indicate that the simple model discussed above
allows us to reproduce the amplification effect observed
experimentally. In the following, it is applied to study the
influence of the permeability $K$ of the grain packing on the
amplification process.

\begin{figure}[t!]
\begin{psfrags}
\psfrag{v}{\small $|v_{bf}|$}
\includegraphics[width=8cm]{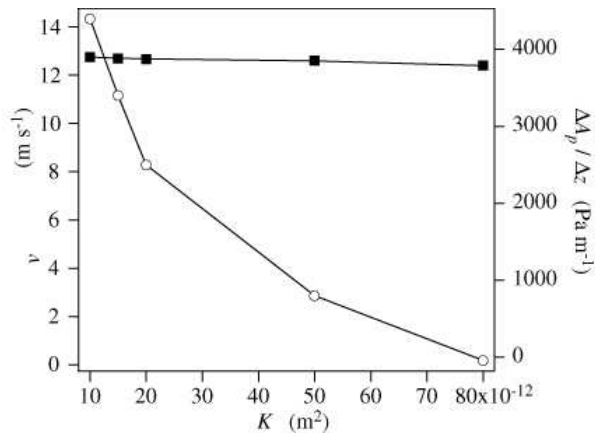}
\end{psfrags}
\caption{Variation of the amplification parameter $\Delta
A_p/\Delta z$ $(\circ)$ and of the absolute front velocity
$|v_{bf}|$ ($\blacksquare$) with the permeability $K$ for
$J$\,=\,450.} \label{fig:kinfluence}
\end{figure}

The permeability influences the dynamics of the system through the
diffusion coefficient $D$ which is proportional to $K$ and appears
in Eq.~(\ref{eq:airconservation3}). The dependance of $|v_{bf}|$
and the amplification parameter $\Delta A_{p}/\Delta z$ on $K$ are
displayed in Fig.~\ref{fig:kinfluence}. The range of
permeabilities investigated is between 0.5 and 4 times the value
$K$\,=\,20\,10$^{-12}$\,m$^2$ assumed in
Fig.~\ref{fig:Jinfluence}.
The front velocity $v_{bf}$ is first almost independent of $K$
while $\Delta A_{p}/\Delta z$ sharply decreases with $K$ and
vanishes for $K\simeq$\,80\,10$^{-12}$\,m$^2$. Note that reducing
$c_{max}$ from 0.63 to 0.61 assuming the influence of wall effects
increases the permeability $K$ from 22 to 27 10$^{-12}$\,m$^2$ and
only reduces the amplification parameter by roughly 15\%.

The above results demonstrate that the diffusion term proportional
to $D$ (and therefore to $K$) in Eq.~(\ref{eq:airconservation3})
is purely dissipative and acts only to enhance the attenuation of
the decompaction wave. This results from the fact that, in high
permeability packing, pressure variations induced by the
decompaction quickly diffuse away from the front. On the contrary,
in low permeability systems, these pressure variations remain
localized: their gradients accelerate nearby grains and so sustain
and amplify the motion. Intermittency effects can therefore be
expected to be stronger for flowing grains of smaller diameter.
\section{Conclusion}
An experimental study has allowed us to analyze the dynamical
properties of periodic intermittent vertical compact flows in a
tube. Our experiments involved simultaneous measurements of local
pressure and particle fraction fields, combined with
spatio-temporal diagrams.

A key observation is the amplification of the particle fraction
and pressure variations as the decompaction front marking the
onset of the flow propagates upwards. This propagation takes place
at a much higher velocity ($|v_{df}|>$\,10\,m\,s$^{-1}$) than that
of the grains (a few 10$^{-1}$\,m\,s$^{-1}$). The velocity
$|v_{bf}|$ of the blockage front is still faster than $|v_{df}|$:
it increases with the flow rate and ranges between 20 and
50\,m\,s$^{-1}$.

The characteristic grain velocity during the flow phase increases
with height above the outlet: this variation is associated with a
decrease with height of the duration $T_f$ of the flow phase as
required by the mass conservation of the grains. Close to the top
of the flow tube, $T_{f}$ represents only 10\% of the total flow
period $T$. The grain velocities and amplitude of the pressure
fluctuations are sufficiently high in this zone that flow from the
feeding hopper gets blocked and a transient low particle fraction
region appears briefly at the top of the tube.

A characteristic feature of these flows is the small amplitude of
the particle fraction variations (0.02--0.03 at most); this is
much smaller than variations observed in the decompaction of a
granular column with an open bottom end or a flow in the density
wave regime. It is noteworthy that these small particle fraction
variations nevertheless induce large pressure drops (up to
3000\,Pa at the top of the tube). The decompaction process is
indeed fast enough for air to undergo an isentropic expansion:
pressure variations closely mirror the variations of the local
particle fraction. Diffusion of air through the grain packing is
too slow to influence strongly that phase; however, it plays an
important role in the longer static phase by allowing the decay of
the pressure towards equilibrium before flow is reinitiated. The
characteristic time for air diffusion through the packing is of
the order of the total period of the intermittent flow and may be
a key factor determining its value.

The amplification effect can be reproduced by a 1D model assuming
an effective driving force on the grains that increases with the
deviation of the particle fraction from the static value. This
numerical model also indicates that the permeability $K$ of the
packing is a very important factor. The amplification effect (and
perhaps also the intermittent flows) should be observable only if
$K$ is sufficiently low that pressure perturbations do not diffuse
away too quickly.

Another important parameter is the relative humidity of the
surrounding air which influence strongly the interaction forces
both between the grains and between the grains and the tube walls.
Preliminary results indicate that the intermittent flow regime is
rarely observed at relative humidities of the order of 40\% or
below, at least with the beads used in the present
study. Systematic studies in a broad range of relative humidities
should greatly help understand better the interaction forces which
determine the characteristics of these intermittent flows.

\section*{ACKNOWLEDGEMENTS}
We wish to thank B. Perrin and E. J. Hinch for helpful suggestions
and discussions during this work. We thank D. Gobin and Ch.
Ruyer-Quil for their help for the numerical simulations and G.
Chauvin, Ch. Saurine and R. Pidoux for the realization of the
experimental setup. We are grateful to J. W. M. Bush and Ph.
Gondret for a thoughtful reading of the manuscript. We finally
wish to thank T. Raafat and V. Terminassian for their
participation to early experiments.

\end{document}